\documentclass[aps,prd,showpacs,preprintnumbers]{revtex4}
\usepackage{amsmath,epsfig}
\usepackage{subfigure}
\usepackage{amssymb,amsfonts}
\usepackage{latexsym}
\usepackage{epsfig}
\usepackage{amssymb}
\usepackage{color}
\usepackage{bm}
\usepackage{dsfont}

\newcommand{\be}{\begin{eqnarray}}
\newcommand{\ee}{\end{eqnarray}}

\newcommand{\bea}{\begin{eqnarray}}
\newcommand{\eea}{\end{eqnarray}}

\begin{document}
	
	
\title{Holographic Technicolor Model and Dark Matter} 
\author{Yidian Chen $^{a}$ }
\thanks{chenyd@mail.ihep.ac.cn}
\author{Xiao-Jun Bi$^{a,b}$}
\thanks{bixj@mail.ihep.ac.cn}
\author{Mei Huang$^{c}$}
\thanks{huangmei@ucas.ac.cn}
\affiliation{$^{a}$  Institute of High Energy Physics, Chinese Academy of Sciences, Beijing 100049, P.R. China}
\affiliation{$^{b}$  School of Physics, University of Chinese Academy of Sciences, Beijing 100049, P.R. China}
\affiliation{$^{c}$  School of Nuclear Science and Technology, University of Chinese Academy of Sciences, Beijing 100049, P.R. China}

\date{\today}
	
\begin{abstract}
We investigate the strongly coupled minimal walking technicolor model (MWT) in the framework of a bottom-up holographic model, where the global $SU(4)$ symmetry breaks to $SO(4)$ subgroup. In the holographic model, we found that 125GeV composite Higgs particles and small Peskin-Takeuchi $S$ parameter can be achieved simultaneously. In addition, the model predicts a large number of particles at the TeV scale, including dark matter candidate Technicolor Interacting Massive Particles (TIMPs). If we consider the dark matter nuclear spin-independent cross-section in the range of $10^{-45}\sim 10 ^ {-48} $cm$^2$, which can be detected by future experiments, the mass range of TIMPs predicted by the holographic technicolor model is 2 $\sim$ 4 TeV.
\end{abstract}

\maketitle

\section{Introduction}	
In 2012, the Higgs boson predicted by the standard model(SM) was discovered by LHC, opening a new era of particle physics\cite{Chatrchyan:2012xdj}.
Although the SM has been very successful, many issues have yet to be resolved, including hierarchy problem and the absence of dark matter particles.
The radiation correction of the Higgs boson requires a huge fine-tuning, which is called a hierarchy problem, meaning that the SM cannot describe physics at higher energy scale, it is just an effective theory of new physics at low energy scale.
	The particles in the SM do not match the properties of dark matter, which means that the new physics contains unknown particles, including dark matter particles.
	From the above two questions, it can be inferred that the new physics contains a new mechanism to solve the hierarchy problem, and to introduce new particles.
	The supersymmetry theory, introducing the symmetry of bosons and fermions, is one of the solutions to these problems. It solves the hierarchy problem and contains possible candidates for dark matter particles.
	In addition, if the Higgs boson is considered as a composite particle, that is, the dynamical electroweak symmetry breaking is introduced, the above problems can also be solved.
	
	
	Since the Higgs boson is an elementary scalar particle, its radiation correction requires a huge fine-tuning.
	As we all know, other spontaneous symmetry breaking in nature comes from the condensation of composite operators.
	Therefore, one solution is tantamount to treat the Higgs boson as a composite particle derived from the new strongly coupled technicolor condensation.
	Therefore, the electroweak part of SM is the effective field theory, and when the energy scale reaches $\Lambda_{TC}$, the details of the new interaction will be revealed.
	The SM does not explain the origin of the spontaneous electroweak symmetry breaking, and technicolor as an alternative idea can avoid the hierarchy problem without introducing the elementary scalar field.
	Technicolor is a new strongly coupled interaction similar to QCD, but it is on the electroweak energy scale\cite{Weinberg:1975gm,Susskind:1978ms}.
	Analogous to Cooper pairs in superconductors, $W$ and $Z$ gauge bosons are obtained by vacuum condensation of techniquarks $\left\langle \bar{Q}_{TC}Q_{TC}\right\rangle $.
	Since there is no elementary Higgs boson, the Yukawa coupling terms in SM are replaced by effective four-fermion interactions, which come from extended technicolor interactions(ETC)\cite{Eichten:1979ah,Dimopoulos:1979es}.
	The flavor-changing neutral currents(PCNC) problem is caused by four-Fermion interactions, which are resolved by walking technicolor(WTC)\cite{Holdom:1984sk,Akiba:1985rr,Yamawaki:1985zg,Bando:1986bg,Appelquist:1986an,Bando:1987we,Appelquist:1987fc}.
	The walking dynamics can avoid PCNC problems by considering a large anomalous dimension $\gamma_m\simeq1$ and can also reduce Peskin-Takeuchi S parameter\cite{Appelquist:1991is,Sundrum:1991rf,Appelquist:1998xf,Harada:2005ru,Kurachi:2006mu,Kurachi:2007at}.
	
	The simplest theory that includes walking dynamics is the Minimal Technicolor Model (MWT), which is $SU(2)$ gauge theory and has two adjoint techniquarks\cite{Sannino:2004qp}.
	In order to avoid the Witten topology anomaly, the model also introduces a new weakly charged fermionic doublet\cite{Dietrich:2005jn}.
	The MWT has $SU(4)$ global symmetry, which breaks into $SO(4)$ symmetry driven by techniquark condensation  $\left\langle Q_i^\alpha Q_j^\beta \epsilon_{\alpha\beta} E^{ij} \right\rangle $.
	The electroweak gauge group is obtained by gauging $SU(2)_L\times SU(2)_R\times U(1)_V$ which is subgroup of the $SU(4)$.
	The $SU(2)_L$ generates the weak gauge group $SU(2)_L$, and the subgroup of $SU(2)_R\times U(1)_V$ generates $U(1)_Y$.
	Techniquark condensation breaks the global $SU(4)$ group to the $SO(4)$, which also drives the gauge $SU(2)_L\times U(1)_Y$ group breaks to $U(1)_Q$.
	The global $SO(4)$ symmetry after breaking is the custodial symmetry of SM.
	The MWT model contains nine pseudo-Goldstone bosons, three of which become the longitudinal part of the $W$ and $Z$ gauge bosons.
	The Higgs boson of SM corresponds to the composite scalar particle in the MWT model.
	The MWT model can not only replace the Higgs part of SM, but also predict the possibility of a strong first-order electroweak phase transition(EWPT)\cite{Kikukawa:2007zk,Cline:2008hr,Jarvinen:2009pk,Jarvinen:2009wr}, and further predict the existence of stochastic gravitational waves generated during the cosmic EWPT period\cite{Jarvinen:2009mh,Jarvinen:2010ms}.
	
	The MWT model contains a wealth of particles beyond SM, including dark matter candidate particles named Technicolor
	Interacting Massive Particles (TIMPs)\cite{Nussinov:1985xr,Chivukula:1989qb,Barr:1990ca,Gudnason:2006yj,Gudnason:2006ug,Kainulainen:2006wq,Khlopov:2007ic,Kouvaris:2007iq,Ryttov:2008xe,Kouvaris:2008hc,Foadi:2008qv,Khlopov:2008ty,Frandsen:2009mi}.
	The simplest of these is the lightest technibaryon with a conservation technibaryon number.
	Similar to protons, the life of such dark matter is very long, and the operators of violating technibaryon number are depressed by the Grand Unified Theories scale.
	TIMPs are produced by the sphaleron transitions, which can be ignored as the temperature decreases, above the electroweak energy scale.
	The weak anomaly will violate baryon number $B$ and lepton number $L$, but protect $B-L$. Similarly, it will break baryon number $B$, lepton number $L$ and technibaryon number $TB$\cite{Barr:1990ca,Kaplan:1991ah,Gudnason:2006yj}.
	But it will protect some combination of $B$, $L$ and $TB$, so it can explain the ratio $\Omega_{DM}/\Omega_B\sim 5$.
	
	Since MWT is a strongly coupled gauge theory, it can be studied by AdS/CTF correspondence or Gauge/Gravity duality\cite{Maldacena:1997re,Gubser:1998bc,Witten:1998qj} (see \cite{Aharony:1999ti,Aharony:2002up,Zaffaroni:2005ty,Erdmenger:2007cm} for review).
	In recent years, many properties of strongly coupled QCD theory, such as meson spectra\cite{Erlich:2005qh,DaRold:2005vr,Karch:2006pv,Sakai:2004cn,Kruczenski:2004me,Sakai:2005yt,DaRold:2005mxj,Ghoroku:2005vt,Andreev:2006ct,Forkel:2007cm,Chen:2015zhh,Ballon-Bayona:2017sxa}, phase transitions, and baryon number susceptibilities\cite{DeWolfe:2010he,DeWolfe:2011ts,Yang:2014bqa,Critelli:2017oub,Li:2017ple} have been extensively studied.
	In addition, holographic electroweak models, including holographic technicolor\cite{Hong:2006si,Hirn:2006nt,Piai:2006hy,Carone:2006wj,Nunez:2008wi,Haba:2010hu,Anguelova:2011bc,Matsuzaki:2012xx,Anguelova:2012ka,Elander:2012fk,Chen:2017cyc} and composite Higgs models\cite{Contino:2003ve,Agashe:2004rs,Agashe:2005dk,Croon:2015wba,Espriu:2017mlq}, have also been studied.
	
In this work, we investigate composite Higgs boson and dark matter by using holographic technicolor model. The paper is organized as following:
In Sec.2 we introduce the holographic technicolor model and holographic Yukawa coupling. We calculated the $S$ parameter and dark matter nuclear
cross-section in Sec.3. Finally, a short summary is given in Sec.4.

\section{5D model Lagrangian}
\label{sec:model}

The new strongly coupled interacting can be described as a holographic 5D model according to AdS/CFT duality.
The 5D model contains scalar and vector fields, corresponding to scalar and vector composite operators, respectively.
Among them, the scalar fields $H$ dual to the operator $\left\langle Q_i^\alpha Q_j^\beta \epsilon_{\alpha\beta} E^{ij} \right\rangle $, that is, the technicolor condenstation driving dynamical electroweak symmetry breaking. The vector fields $A^M$ are connected with the techniquark bilinear operator $Q_i^\alpha \sigma_{\alpha\dot{\beta}}^\mu \bar{Q}^{\dot{\beta},j}-\frac{1}{4}\delta_i^j Q_k^\alpha \sigma_{\alpha\dot{\beta}}^\mu \bar{Q}^{\dot{\beta},k}$.
    In addition, the model also includes the dilaton field $\phi(z)=\mu z^2$ which is similar to AdS/QCD models to describe the Regge slope\cite{Karch:2006pv}.

    In the Poincar\'{e} patch, the 5D AdS metric is
    \begin{eqnarray}
    ds^2=g_{MN}dx^M dx^N=\frac{L^2}{z^2}(\eta_{\mu\nu}dx^\mu dx^\nu+d^2z),\quad \eta_{\mu\nu}=\text{diag}(-1,1,1,1).
    \end{eqnarray}
    In general, the AdS radius $L$ is set to $1$.
    The $SU(4)$ invariant action is assumed as
    \begin{eqnarray}\label{5d-action}
    S_5=&-\int d^5x\sqrt{-g}~e^{-\phi(z)}\Bigg\{ \frac{1}{2} {\rm Tr}~ \bigg[(D^MH)^\dagger(D_MH)+m_5^2 H^\dagger H+\lambda\phi H^\dagger H\bigg]\nonumber\\
    &+\frac{1}{4g_5^2}{\rm Tr}~F^{MN}F_{MN}\Bigg\},
    \end{eqnarray}
    with $m_5^2=(\Delta-\gamma_m)(\Delta-\gamma_m-4)$ and $g_5^2=12\pi^2/N_{TC}$.
    The anomalous dimension $\gamma_m$ is set to $1$ on the basis of walking technicolor mechanism and the 5D mass satisfies Breitenlohner-Freedman bound $m_5^2=-4$.
    The scalar field $H$ describing dynamical breaking from $SU(4)$ to $SO(4)$ can be expanded as the nonlinear form:
    \begin{eqnarray}
    H=e^{2i\Pi^a(x,z)T^a}\frac{v(z)+h(x,z)}{2}E,
    \end{eqnarray}
    where
    \begin{eqnarray}
    E=\begin{pmatrix}
    & \mathds{1}_{2\times 2}\\
    \mathds{1}_{2\times 2} &
    \end{pmatrix}.
    \end{eqnarray}
    The composite scalar field $H$ corresponds to the Higgs field in the standard model, and the lowest KK excited state of the scalar field $h$ corresponds to the Higgs boson.
    The field $v$ in the expansion of scalar field $H$ indicates the technicolor condensation, that is, the dynamical electrocweak symmetry breaking.
    Breaking from SU(4) to SO(4), nine Goldstone particles are produced, three of which become the longitudinal part of the $W$ and $Z$ gauge bosons, and the remaining six contain candidates of dark matter particles.
    Holography duals the global symmetry of boundary theory to gauge symmetry of bulk theory.
    Thus the covariant derivative is defined as
    \begin{eqnarray}
    D_MH=\partial_MH-iA_MH-iHA_M^T,
    \end{eqnarray}
    where $T$ represents the transpose of the matrix.
    The $\lambda$ term of action represents the interaction between the dilaton field and the scalar field.
    Since the dilaton field $\phi\to 0$ when $z\to 0$, the behavior of scalar field $v$ does not change in the UV region.
    As we will see in the next section, the scalar field $v$ has tiny changes when $\lambda$ is close to $-4$.

    The strength tensor of vector fields is
    \begin{eqnarray} F_{MN}=(\partial_MA_N^A-\partial_NA_M^A-i\left[A_M^A,A_N^A\right])T^A,
    \end{eqnarray}
    where the generators $T^A$ indicate both broken ($T^a$) and unbroken ($S^i$) case.
    The representation of the generators can be referred to Ref. \cite{Appelquist:1999dq}.
    It is worth noting that the vector fields $A$ are not the SM electroweak gauge fields $W$ or $Z$.
    But it will mix with electroweak fields when the $W$ and $Z$ are introduced.

\subsection{Scalar Vacuum Expectation Value}
The scalar vacuum expectation value in Eq.(\ref{5d-action}) can be obtained from the following equation
\begin{eqnarray}
    -\frac{z^3}{e^{-\phi(z)}}\partial_z \frac{e^{-\phi}}{z^3}\partial_z v(z)+\frac{m_5^2+\lambda\phi}{z^2}v(z)=0.
\end{eqnarray}
In order to obtain the mass of the Higgs boson, $\lambda$ is considered to be close to $-4$. Considering the behavior of $v(z)$ when $z$ is large, the equation can be approximated as
    \begin{eqnarray}
    -v''(z)+(2\mu z)v'(z)+\lambda\mu v(z)=0.
    \end{eqnarray}
And then, $v$ tends to be $v\sim z^2$. This is similar to the solution of $v(z)$ in the hard-wall model with $m_5^2=-4$. Since the behavior of the scalar field $v$ in the UV region is not changed, the approximation of $v=Mz^2$ can be considered.

\begin{figure}[htbp]
    \centering
    \includegraphics[height=6.0cm,width=9.5cm]{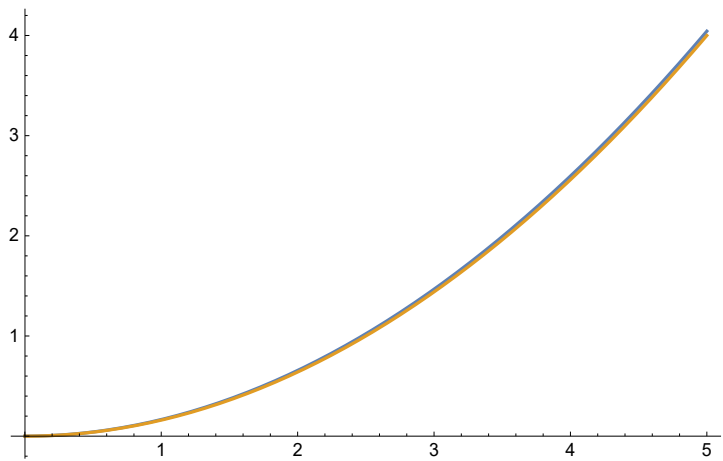}
    \caption{The difference between the numerical and the approximate solution of $v(z)$. The blue and orange lines are numerical and approximate results, respectively.}
    \label{fig:dif-vac}
    \end{figure}

Numerical solution indicates that $v(z)=M z^2$ is a good approximation. The UV boundary condition $v\to z^2$ is set when solving the numerical solution.
As shown in Figure \ref{fig:dif-vac}, the difference between the numerical and the approximate solution of $v(z)$ is very small.
Further calculations find that the approximation has little effect on other numerical results. So in the following we only consider the approximation $v(z)=M z^2$ in order to get more analytical results.

\subsection{Scalar field}

The EOM for the scalar field $h(x,z)$ is
    \begin{eqnarray}\label{m-s}
    -\frac{z^3}{e^{-\phi(z)}}\partial_z \frac{e^{-\phi}}{z^3}\partial_z h -q^2 h
    +\frac{1}{z^2}(m_5^2+\lambda\phi)h=0.
    \end{eqnarray}
The solution is
    \begin{eqnarray}\label{sol-s}
    h(q, z)=C_1(q)e^{\mu z^2}z^2U(-\frac{q^2+\lambda\mu}{4\mu},1,-\mu z^2)+C_2(q)e^{\mu z^2}z^2L(\frac{q^2+\lambda\mu}{4\mu},-\mu z^2).
    \end{eqnarray}

The first term represents the bulk-to-boundary propagator and the second term gives us the KK towers of scalar field. Normalizable solution is given as
\begin{eqnarray}
    h_n(z)=\sqrt{\frac{2}{\mu}}\mu z^2L_n(\mu z^2),\quad M_h^2(n)=4\mu(n+1+\frac{\lambda}{4}), \quad n=0,1,2,...
\end{eqnarray}
The orthogonality relation is
    \begin{eqnarray}
    \int_0^\infty dz \frac{e^{-\phi(z)}}{z^3}h_n(z)h_m(z)=\delta_{nm}.
    \end{eqnarray}
It can be found from the above equation that when $\lambda$ approaches 0, the KK excited state of scalar field and unbroken vector fields are degenerate.
This means that $\lambda$ introduces the splitting of mass of scalar field and unbroken vector fields.

In order to obtain the Higgs boson of SM, $\lambda$ is set to $\lambda=-4+\frac{1}{64\mu}$. Then the mass of the lowest KK excited state of the scalar field is 125 GeV. If we consider that other particles are in the TeV scale, then Regge slope parameter $\mu$ should be greater than $1/4$.
This is consistent with the approximation that $\lambda$ approaches $-4$.

\subsection{Unbroken Vector fields}

Expanding the action in Eq.(\ref{5d-action}), the EOM of transverse part of unbroken vector fields are obtained as
 \begin{eqnarray}\label{eom-v}
    -\frac{z}{e^{-\phi(z)}}\partial_z(\frac{e^{-\phi(z)}}{z}\partial_z A_\mu^i(q,z))+q^2 A_\mu^i(q,z)=0,
 \end{eqnarray}
where $i=1,...,6$ and $A_z=0$ gauge is considered. And $A_\mu^i(q,z)$ are the 4D Fourier transform of $A_\mu^i(x,z)=\int d^4q e^{iqx}A_\mu^i(q,z)$.

According to holography, the fields $A_\mu^a(q,z)$ can be written as
    \begin{eqnarray}
    A_\mu^i(q,z)=V(q,z){\mathcal V}_\mu^i(q),\quad\ V(q,\epsilon)=1.
\end{eqnarray}
The exact solution is
    \begin{eqnarray}\label{sol-v}
    V(q,z)=C_1(q)U(\frac{q^2}{4\mu},0,\mu z^2)+C_2(q)L(-\frac{q^2}{4\mu},-1,\mu z^2),
    \end{eqnarray}
where $U$ is the Tricomi's confluent hypergeometric function and $L$ is generalised Laguerre polynomials.

The first term represents the bulk-to-boundary propagator and the second term gives us the KK towers of vector fields.
Normalizable solutions are given as
\begin{eqnarray}\label{m-v}
    V_n (z)=\mu z^2\sqrt\frac{2}{n+1} L_n^1(\mu z^2),\quad \
    M_V^2(n)=4\mu (n+1),\quad n=0,1,2,...
\end{eqnarray}
The $V_n(z)$ fulfil the following orthogonality relation:
\begin{eqnarray}
\int_0^\infty dz \frac{e^{-\phi(z)}}{z}V_n(z)V_m(z)=\delta_{nm}.
\end{eqnarray}
This result is similar to the result of AdS/QCD, the masses of vector particles are determined by $\mu$.

    \subsection{Broken Vector fields}
    The EOM of transverse part of broken vector fields are
    \begin{eqnarray}\label{eom-a}
    \left(-\frac{z}{e^{-\phi(z)}}\partial_z \frac{e^{-\phi(z)}}z\partial_z A_\mu^a-q^2 A_\mu^a-\frac{g_5^2 v(z)^2}{z^2}A_\mu^a\right)_\bot=0.
    \end{eqnarray}
    where $a=1,...,6$ and $A_z=0$ gauge is considered.

    In order to get an analytical solution we have to use the approximation that $v(z)= M z^2$.
    Then, the solution is
    \begin{eqnarray}\label{sol-a}
    A(q,z)=C_1(q)e^{\frac{(\mu-\tilde{\mu})z^2}{2}}U(\frac{q^2}{4\tilde{\mu}},0,\tilde{\mu}z^2)+C_2(q)e^{\frac{(\mu-\tilde{\mu})z^2}{2}}L(-\frac{q^2}{4\tilde{\mu}},-1,\tilde{\mu}z^2),
    \end{eqnarray}
    where $\tilde{\mu}=\sqrt{g_5^2M^2+\mu^2}$.

    The first term represents the bulk-to-boundary propagator and the second term gives us the KK towers of vector fields.
    Normalizable solutions are given as
    \begin{eqnarray}\label{m-a}
    A_n (z)=\sqrt{\frac{2}{n+1}}e^{\frac{(\mu-\tilde{\mu})z^2}{2}}\tilde{\mu} z^2L_n^1(\tilde{\mu}z^2), \quad M_A^2(n)=4\tilde{\mu}(n+1) \quad  n=0,1,2,....
    \end{eqnarray}
    The orthogonality relation is
    \begin{eqnarray}
    \int_0^\infty dz \frac{e^{-\phi(z)}}{z}A_n(z)A_m(z)=\delta_{nm},
    \end{eqnarray}

    We observe that the Regge trajectory of broken vector fields are similar to the unbroken vector fields, but the slope is larger.
    So the broken vactor states are heavier than their unbroken counterparts.
    It is worth noting that this conclusion is only valid when the approximation $v=Mz^2$ is applied.
    This means that $\lambda$ must approach $-4$, which is consistent with previous result.
    If the numerical solution is performed, the numerical results of the vector particle spectrum are not much different from the analytical results, indicating that the approximation $v=Mz^2$ is suitable.

\subsection{Goldstone bosons and dark matter particles}
    The EOM of Goldstone bosons are
    \begin{eqnarray}
    \partial_z \frac{e^{-\phi(z)}}{z}\partial_z\varphi^a+\frac{e^{-\phi(z)}g_5^2v(z)^2}{z^3}(\Pi^a-\varphi^a)&=&0,\\
    \frac{g_5^2v(z)^2}{z^2}\partial_z\Pi^a+q^2\partial_z\varphi^a&=&0,
    \end{eqnarray}
    where $a=1,2,,,9$. By eliminating the $\varphi$ in the above coupled equations, we can get the following equation
    \begin{eqnarray}
    -\partial_z \frac{z^3}{e^{-\phi(z)}v(z)^2}\partial_z\frac{e^{-\phi(z)}v(z)^2}{z^3}\Pi'^a+q^2\Pi'^a+\frac{e^{-\phi(z)}g_5^2v(z)^2}{z^2}\Pi'^a=0,
    \end{eqnarray}
where $\Pi'^a$ is the derivative of $\Pi^a$. In order to get an analytical solution we have to use the approximation that $v(z)\sim M z^2$.
Then, the $\Pi'^a$ solution is
    \begin{eqnarray}\label{sol-pi}
    \Pi'^a(q,z)=\frac{1}{Mz}~e^{\frac{(\mu-\tilde{\mu})z^2}{2}}\Big[C_1(q)U(\frac{q^2}{4\tilde{\mu}},0,\tilde{\mu}z^2)+C_2(q)L(-\frac{q^2}{4\tilde{\mu}},-1,\tilde{\mu}z^2)\Big].
    \end{eqnarray}

The first term represents the bulk-to-boundary propagator and the second term gives us the KK towers of vector fields. So the mass spectra are
    \begin{eqnarray}
    M_{\Pi}^2(n)=4\tilde{\mu}(n+1) \quad  n=0,1,2,....
    \end{eqnarray}
It can be observed that the pseudoscalar fields and the broken vector fields are degenerate on the approximation of $v(z)= Mz^2$.

 In this model, there are nine Goldstone particles, three of which become the longitudinal parts of the $W$ and $Z$ bosons.
 The remaining six Goldstone bosons include $UU$, $DD$, and $UD$ technibaryons\cite{Gudnason:2006yj}, and their electric charges are $t+1$, $t-1$, and $t$, respectively, where $t$ depends on the representation. Without loss of generality, let $t=1$, then $UU$ is dark matter candidate TIMP.
 For convenience, we mark $\Pi(z)$ corresponding to $UU$ the dark matter particles $\chi(z)$.

\subsection{Interaction between quarks and dark matter particles}

In this section, the SM gauge bosons and quarks Yukawa coupling are introduced to the holographic model.
Therefore, the mass of the $W$ and $Z$ bosons and the interaction between quark and dark matter particles can be obtained.

Modifying the covariant derivatives, the SM gauge field is naturally introduced into the holographic model.
According to the principle of gauge invariance, the covariant derivative has the following form
    \begin{eqnarray}
    D_MH \to \partial_MH-iA_MH-iHA_M^T-iG_MH-iHG_M^T,
    \end{eqnarray}
where
\begin{eqnarray}
    G_M=W_M^\alpha L^\alpha+Z_MY,\\
    L^\alpha=\frac{S^\alpha+T^\alpha}{\sqrt{2}}, \quad Y=\frac{S^3-T^3}{\sqrt{2}}+\sqrt{2}yS^4,
\end{eqnarray}
with $\alpha=1,2,3$. The $y$ in the above equation depends on the representation, and different $y$ correspond to different dark matter particles.
In the holographic model, the specific value of $y$ has no effect on the following results. $W$ and $Z$ are SM gauge fields, and they are assumed to be independent of the fifth dimensional coordinate $z$. Since the techniquark condensation breaks the electroweak symmetry, $W$ obtains the mass
    \begin{eqnarray}
    m_W^2=\int dz ~e^{-\phi}\frac{gv^2}{8z^3}.
    \end{eqnarray}
If vacuum expectation value on the approximation of $v(z)= Mz^2$, the mass of $W$ boson is
    \begin{eqnarray}
    m_W=\frac{gM}{2\sqrt{2\mu}},
    \end{eqnarray}
where $g$ is the $SU(2)$ gauge coupling in the SM. From the above equation we can get the techni-pion decay constant as
    \begin{eqnarray}\label{pi-decay}
    F_\Pi=\frac{M}{\sqrt{2\mu}}.
    \end{eqnarray}

In the standard model, quarks and leptons obtain masses through Yukawa coupling. Since there is no elementary scalar field in the technicolor model, it is necessary to introduce coupling terms between the composite scalar field and the quarks. In order to extend SU(4) symmetry to quarks, we introduce the following vector\cite{Foadi:2007ue}
\begin{eqnarray}
    q^j=\begin{pmatrix}
    u_L^j\\
    d_L^j\\
    -i\sigma^2u_R^{j~*}\\
    -i\sigma^2d_R^{j~*}
    \end{pmatrix},
\end{eqnarray}
where $j$ is generation index. Then, yukawa coupling term is introduced into the holographic model
    \begin{eqnarray}\label{action-y}
    \mathcal{L}_Y&=&-y_u^{ij}q^{iT}P_uM^*P_uq^{j}-y_d^{ij}q^{iT}P_dM^*P_dq^j+h.c.,\\
    P_u&=&p_u(z)\begin{pmatrix}
    1_{2\times 2}& \\
    & \frac{1+\sigma^3}{2}
    \end{pmatrix},
    P_d=p_d(z)\begin{pmatrix}
    1_{2\times 2}& \\
    & \frac{1-\sigma^3}{2}
    \end{pmatrix},
    \end{eqnarray}
where $P_u$ and $P_d$ represent the projection operators of $SU(2)_R$ breaking to $U(1)_R$. Since the functions $p_u(z)$ and $p_d(z)$ come from ETC interactions, their forms are related to the details of the ETC, so they are assumed to be $p_{u/d}\sim z^2$. From action (\ref{action-y}), the yukawa coupling term of quarks and the interaction between quarks and dark matter particles can be given as
\begin{eqnarray}\label{action-qd}
    \Delta S=-\int d^5x \frac{e^{-\phi}}{z^5}v(z)\sum_{f=1}^{6}p_f(z)^2\Big(y_f\bar{q}_fq_f-\frac{y_f}{2}\bar{q}_fq_f\chi^\dagger(z)\chi(z)\Big),
\end{eqnarray}
where the dark matter particles $\chi$ are the $UU$ components of Goldstone particles $\Pi(z)$ and the pecial representation of $\chi$ depends on the value of $y$ \cite{Gudnason:2006ug}.
The specific value of $y$ has little effect on the discussion of this article, so we will not discuss it in detail.
It is worth noting that the dark matter particles depend on the fifth dimensional coordinate $z$, whereas the quarks are independent of $z$.

\section{Results}
\label{sec:result}

\subsection{Correlation Functions and $S$ Parameter}

According to the AdS/CFT duality, two-point correlation function can be obtained as the second derivative of the action with respect to the source.
    So the correlation function can be written as
    \begin{eqnarray}
    \langle \mathcal{O}(x_1)\mathcal{O}(x_2)\rangle= \left. \frac{\delta^2}{\delta \phi_0[x_1]\delta \phi_0[x_2]}e^{-S_{\rm sugra}[\phi[\phi_0]]}\right|_{\phi_0=0}.
    \end{eqnarray}
    If the source is the vector current operator, the correlator has the following form
    \begin{eqnarray}
    \int d^4x e^{iqx}\langle J_\mu^a(x)J_\nu^b(0)\rangle=\delta^{ab}(\frac{q_\mu q_\nu}{q^2}-g_{\mu\nu})\Pi_V(q^2).
    \end{eqnarray}
    Considering the on-shell action (\ref{5d-action}), then $\Pi_V$ is
    \begin{eqnarray}
    \Pi_V(q^2)= \frac{1}{g_5^2}\left.\left[\frac{e^{-\phi(z)}V(q,z)\partial_z V(q,z)}{z}\right]\right|_{z=\epsilon}.
    \end{eqnarray}
    Similar to the unbroken case, the broken vector current correlator is given by
    \begin{eqnarray}
    \Pi_A(q^2)= \frac{1}{g_5^2}\left.\left[\frac{e^{-\phi(z)}A(q,z)\partial_z A(q,z)}{z}\right]\right|_{z=\epsilon}.
    \end{eqnarray}

    From holography, the KK part of $V$ and $A$ has little effect on the correlator, and only the bulk-to-boundary propagator is significant.
    Substituting the propagator from Eqn.(\ref{sol-v}), the unbroken vector correlator is given as
    \begin{eqnarray}
    \Pi_V(q^2)=\frac{q^2}{2g_5^2}(2\gamma_E+{\rm ln}~\mu z^2+\psi(1+\frac{q^2}{4\mu})),
    \end{eqnarray}
    with $\gamma_E$ is the Euler constant and $\psi$ is the digamma function.
    Here we use the boundary condition of $V(q,\epsilon)=1$.

    Similarly, broken vector correlator can be obtained from Eqn.(\ref{sol-a})
    \begin{eqnarray}
    \Pi_A(q^2)=\frac{\mu-\tilde{\mu}}{g_5^2}+\frac{q^2}{2g_5^2}(2\gamma_E+{\rm ln}~\tilde{\mu} z^2+\psi(1+\frac{q^2}{4\tilde{\mu}})).
    \end{eqnarray}
    Again, we use the boundary condition of $A(q,\epsilon)=0$.
    It can be observed that when $\mu=\tilde{\mu}$, the correlator of the unbroken and broken vector are consistent.
    In this case, technicolor condensate $M=0$ and the $SU(4)$ symmetry is unbroken.

    The $S$ and $T$ of Peskin-Takeuchi parameters are important for the exploration of new physics.
    Due to the existence of custodial symmetry, T disappears in this model, so only S parameters are considered.
    $S$ parameter can be obtained by unbroken and broken vector correlators\cite{Haba:2008nz}
    \begin{eqnarray}\label{S-para}
    S=\left. -4\pi \frac{d}{dq^2}(\Pi_V-\Pi_A)\right|_{q^2\to 0}=\frac{2\pi}{g_5^2}{\rm ln}~\frac{\tilde{\mu}}{\mu}.
    \end{eqnarray}
    According to the definition of $\tilde{\mu}$, $\tilde{\mu}$ is greater than $\mu$, and thus $S$ is positive.
    We can also get the decay constant of techni-pion
    \begin{eqnarray}\label{pi-decay-2}
    F_\Pi^2=\Pi_V(0)-\Pi_A(0)=\frac{\tilde{\mu}-\mu}{g_5^2}.
    \end{eqnarray}
    We can observe that the results of (\ref{pi-decay}) and (\ref{pi-decay-2}) seem to be inconsistent.
    However, if we consider the approximation of $\mu \gg g_5M$, the results of (\ref{pi-decay-2}) will become (\ref{pi-decay}).
    On the approximation, the S parameter will become
    \begin{eqnarray}
    S\simeq \frac{2\pi F_\Pi^2}{\mu},
    \end{eqnarray}
    and it is consistent with the strong dynamics $S\approx 4\pi F_\Pi^2(M_V^{-2}+M_A^{-2})$\cite{Contino:2010rs}.

\begin{figure}
    \centering
    \includegraphics[height=6.0cm,width=9.5cm]{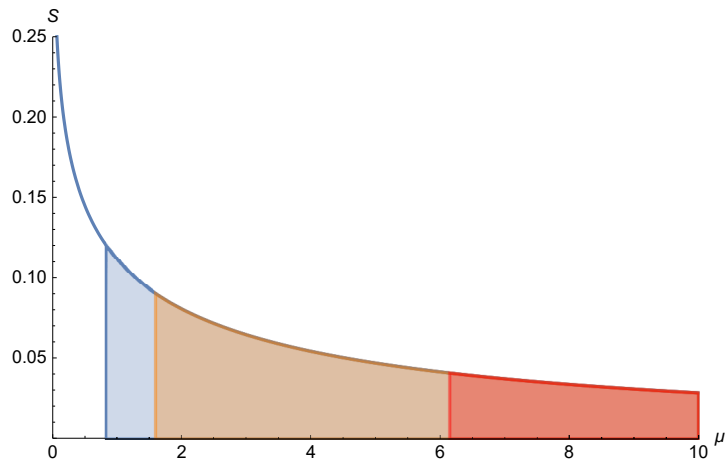}
    \caption{The range of values of S and $\mu$ when different conditions are satisfied. The blue region indicates that $-0.08\leq S \leq 0.12$; the orange region indicates that $-0.05\leq S \leq 0.09$; the red region indicates that Eqn.(\ref{pi-decay}) and (\ref{pi-decay-2}) are consistent.}
    \label{fig:S-mu}
    \end{figure}

    In the holographic model, if the Yukawa term is not included, it has 4 parameters: $N_{TC}$, $M$, $\mu$ and $\lambda$.
    Since the color $N_{TC}$ of technicolor has little effect on the result, it is fixed to $2$.
    Fitting the mass of Higgs boson and $W$ boson, the model has only one free parameter $\mu$, ie, the Regge slope of the particles.
    As can be seen from the Eqn.(\ref{S-para}) and Fig.(\ref{fig:S-mu}), the $S$ parameter monotonically decreases as the $\mu$ increases.
    From the PDG\cite{Tanabashi:2018oca}, $S$ parameter is required to be within the range: $-0.08\leq S \leq 0.12$ or $-0.05\leq S \leq 0.09$($U=0$ is fixed).
    Therefore, it can be seen from the Eqn.(\ref{S-para}) that $\mu$ must be satisfied $\mu\gtrsim 0.83$ or $\mu\gtrsim 1.6$.
    The $F_\Pi$ of the Eqn.(\ref{pi-decay-2}) monotonically increases as the $\mu$ becomes larger.
    If the difference between (\ref{pi-decay-2}) and (\ref{pi-decay}) is less than 10\%, then $\mu$ must satisfy $\mu \gtrsim 6.15$.

    \subsection{Dark Matter Direct Detection}

In this section, we will consider dark matter particles in the holographic model.
In the model, dark matter are pseudo Goldstone particles produced by spontaneous symmetry breaking.
The dark matter particles have a techni-baryon number, so during the cosmic electroweak phase transition, enough dark matter are produced by the sphaleron process\cite{Gudnason:2006ug}. Through the sphaleron process, baryon energy density can be linked with dark matter density.
And when the mass of the dark matter is about $2.2$TeV, the relic density can be obtained\cite{Sannino:2009za}.

In the holographic model, dark matter particles interact with quarks through the Yukawa coupling term.
    Calculating the cross section of dark matter and nucleus, the parameter space of the model can be constrained.
    Due to the dark matter relic density, we mainly focus on the region of TeV scale.
    It can be seen from (\ref{action-qd}) that the Yukawa coupling $y_f$ is adjusted to fit the quark mass, and the effective coupling constants of the interaction between the dark matter particles and quarks can be obtained.
    Thus, the quark mass can be given as
    \begin{eqnarray}
    m_f=y_f\int_{0}^{\infty}dz\frac{e^{-\phi}}{z^5}v(z)p_f(z)^2,
    \end{eqnarray}
    where $f$ is flavor of quarks.
    It is worth noting that the above equation contains the unknown function $p_f$, which is derived from the ETC interaction.
    We assume that its behavior is $p_f=z^2$, ie it has a similar form to the vacuum expectation value $v(z)$.
    Since the coefficient of the function can be absorbed into $y_f$, it is set to $1$.
    And the effective coupling constants $F_f$ is
    \begin{eqnarray}
    F_f=-\frac{y_f}{2}\int_{0}^{\infty}dz\frac{e^{-\phi}}{z^5}v(z)p_f(z)^2\chi^\dagger(z)\chi(z),
    \end{eqnarray}
    where $\chi (z)$ is given by Eqn.(\ref{sol-pi}).
    Eqn.(\ref{sol-pi}) only gives the derivative of $\chi$, and additional boundary condition needs to be added.
    By selecting the boundary condition $\Pi''(z\to\infty)=\Pi'(\epsilon)=\Pi(\epsilon)=0$, the dark matter $\chi$ can be solved and the effective coupling constants can be given.

    The dark matter nucleus cross section can be obtained by the following\cite{Yu:2011by}
    \begin{eqnarray}
    \sigma_{SI}=\frac{m_N^2}{4\pi(M_{DM}+m_N)^2}(\frac{F_N}{\sqrt{2}})^2,
    \end{eqnarray}
    where $M_{DM}$ and $m_N$ are dark matter mass and nucleus mass, respectively, $F_N$ is induced coupling constants of dark matter nucleus interactions.
    $F_N$ and $F_f$ are related by
    \begin{eqnarray}
    F_N=\sum_{f=u,d,s}F_ff_f^N\frac{m_N}{m_f}+\sum_{f=c,b,t}F_ff_Q^N\frac{m_N}{m_f},
    \end{eqnarray}
    with the nucleon form factors $f_u^p=0.020\pm0.004$, $f_d^p=0.026\pm0.005$, $f_s^p=0.118\pm0.062$, $f_u^n=0.014\pm0.003$, $f_d^n=0.036\pm0.008$, $f_s^n=0.118\pm0.062$ and $f_Q^N=\frac{2}{27}(1-f_u^N-f_d^N-f_s^N)$ for heavy quarks\cite{Ellis:2000ds,Alarcon:2011zs}.

\begin{figure}
    	\centering
    	\includegraphics[height=6.0cm,width=9.5cm]{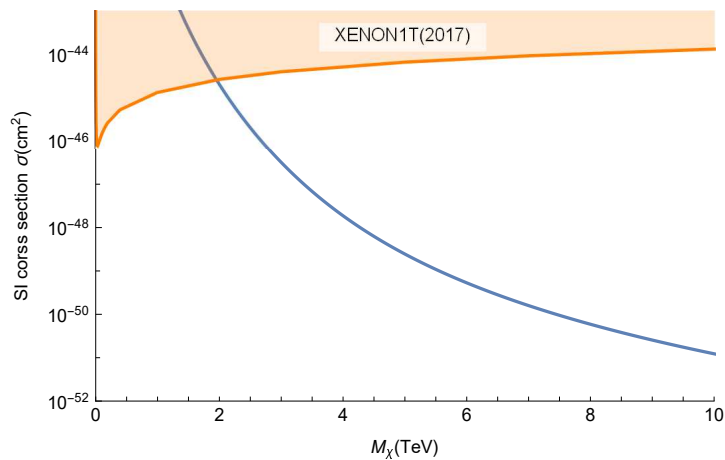}
    	\caption{The spin-independent (SI) dark matter-nucleon cross sections: The blue line is the result of the holographic dark matter particles $\chi$; the orange line is the experimental data of XENON1T\cite{Aprile:2017iyp}.}
    	\label{fig:SI}
    \end{figure}

The dark matter nucleon scattering cross section can be obtained by the effective coupling constant, as shown in Fig.(\ref{fig:SI}).
It can be seen from Fig.(\ref{fig:SI}) that the orange part has been excluded by the XENON1T experiment.
As we can see from Fig.(\ref{fig:SI}), the cross section decreases as the mass of the dark matter increases, and intersects the XENON1T experimental data at approximately $2$TeV. Therefore, the case where the mass is less than $2$TeV has been ruled out by the experiment, ie $\mu\lesssim 0.14$.
Considering the possible range of direct detection for future experiments, we focus on the case where the cross section is $10^{-45}\sim 10^{-48}{\rm cm}^2$. In other words, we pay more attention to the situation that the dark matter mass is at $2\sim4$TeV, corresponding to $1.79\gtrsim\mu\gtrsim 0.14$.
For the case where the mass is greater than $4$TeV, since the dark matter particles are difficult to detect, the constraint on the holographic model is small.

When the dark matter mass is considered to be much larger than the electroweak phase transition temperature, the dark matter mass estimated by the electroweak sphaleron process is about 2TeV\cite{Sannino:2009za}, which is consistent with the lower limit we estimate by direct detection in holographic model. This means that the mass of dark matter in the model that satisfies direct detection can explain the relic density $\Omega_{DM}/\Omega_{B}\sim5$.
For the case where the electroweak phase transition temperature is much larger than the dark matter mass, the dark matter mass estimated by the sphaleron phase transition is about 5TeV\cite{Sannino:2009za}. The upper mass limit calculated in the holographic model means that the phase transition temperature is comparable to the dark matter mass. Heavier dark matter, that is, the larger parameter $\mu$, is associated with higher electroweak phase transition temperature.

Further considering the constraints of the $S$ parameter, the range of the parameter $\mu$ is $1.79\gtrsim\mu \gtrsim 0.83$ or $1.79\gtrsim\mu\gtrsim 1.6$, that is, the dark matter mass is $4$TeV$\gtrsim M_{DM} \gtrsim 3.22$TeV or $4$TeV$\gtrsim M_{DM} \gtrsim 3.88$TeV, respectively.
The constraint of the $S$ parameter requires the model to have heavier dark matter, and the SI section is $10^{-47}\sim10^{-48}$cm$^2$, which implies a higher phase transition temperature in holographic technicolor model. Since the consistency of Eqn.(\ref{pi-decay}) and Eqn.(\ref{pi-decay-2}) requires $\mu$ is greater than $6.15$, which makes the cross section of dark matter and nucleus too small, and therefore this is not in the scope of attention.

In summary, considering the constraints of the relic density, SI cross section, and $S$ parameter, the dark matter mass is about $4$TeV$\gtrsim M_{DM} \gtrsim 3.22$TeV or $4$TeV$\gtrsim M_{DM} \gtrsim 3.88$TeV. In this case, the SI cross section is $10^{-47}\sim10^{-48}$cm$^2$, and the relic density requires that the phase transition temperature be comparable to the dark matter mass.

    \section{Conclusions}
    \label{sec:conclusions}
    In this work, we studied dynamical electroweak symmetry breaking and dark matter by using the gauge/gravity duality.
    We successfully constructed a holographic technicolor model that is dual to the MWT model with $N_{TC}=2$, in which the $W$ and $Z$ bosons obtained masses by technicolor condensation.
    In adition, we calculated the Peskin-Takeuchi S parameters and obtained many particles at the TeV energy scale, including dark matter candidate TIMPs.

    In this holographic model, similar to QCD, the gauge boson obtains mass by technicolor condensation, and the $125$GeV boson, similar to $\sigma$ particle in QCD, is composite Higgs boson.
    If the mass of the Higgs and $W$ bosons is fitted, the holographic model has only one free parameter $\mu$ left.
    $\mu$ describes the Regge slope of technihadrons and determines the mass of technihadrons.
    The $S$ parameter is used to constrain the parameter space of the new physics and its experimental range is $-0.08\leq S \leq 0.12$ or $-0.05\leq S \leq 0.09$($U=0$ is fixed).
    Since in the holographic model, the $S$ parameter decreases as $\mu$ increases, $\mu$ needs to satisfy $\mu\gtrsim 0.83$ or $\mu\gtrsim 1.6$.

    Among the many technihadrons of the holographic model, dark matter candidate particles TIMPs are included.
    By adding the Yukawa coupling term, the holographic model can obtain the effective coupling constant of the dark matter and quarks, and further obtain the spin-independent dark matter nucleon cross section $\sigma_{SI}$.
    We found that the cross section in the model decreases as the mass of the dark matter increases, and the theoretical line intersects the XENON1T experimental line at the dark matter mass of approximately $2$TeV.
    If we are concerned about the range of $10^{-45}\sim 10^{-48}$cm$^2$ that may be detected in future experiments, the dark matter mass is limited to $2\sim4$TeV.
    If both $S$ parameter and dark matter cross section constraints are considered, the mass of dark matter is $3.2\sim 4$TeV or $3.8\sim 4$TeV.

\vskip 0.5cm
{\bf Acknowledgement}
\vskip 0.2cm

Y.D.C. is supported by the NSFC under Grant No.11847232, X.J.B. is supported by the the National Natural Science Foundation of China (Grants No. U1738209 and No. 11851303), and M.H. is supported in part by the NSFC under Grant Nos. 11725523, 11735007, 11261130311 (CRC 110 by DFG and NSFC), Chinese Academy of Sciences under Grant No. XDPB09, and the start-up funding from University of Chinese Academy of Sciences(UCAS).

	
	%
	%
	
	%
	
	
	\bibliography{ref}
	\bibliographystyle{unsrt}
	
\end{document}